\documentclass[epj]{svjour}
\usepackage{graphicx,psfrag}
\usepackage{dcolumn}
\usepackage{bm}
\usepackage{subfigure}
\usepackage{amsmath}
\usepackage{amssymb}
\newcommand{\norm}[1]{{\cal N}_{#1}}
\usepackage{ulem} 
\usepackage[usenames]{color}

\begin{document}
\title{Kaonic production of $\Lambda(1405)$ off deuteron target in chiral dynamics}
\author{D. Jido\inst{1} \and E. Oset\inst{2}
\and T. Sekihara\inst{3}}
%
%
\institute{
Yukawa Institute for Theoretical Physics, 
Kyoto University, Kyoto 606-8502, Japan 
\and 
Departamento de F\'{\i}sica Te\'orica and IFIC,
Centro Mixto Universidad de Valencia-CSIC,
Institutos de Investigaci\'on de Paterna, Aptdo. 22085, 46071 Valencia, Spain
\and
Department of Physics, Graduate School of Science, Kyoto University,
Kyoto, 606-8502, Japan}
\date{Received: date / Revised version: date}
%
\abstract{
The $K^{-}$ induced production of $\Lambda(1405)$ is investigated 
in $K^{-} d \to \pi \Sigma n$ reactions based on coupled-channels 
chiral dynamics, in order to discuss the resonance position of the 
$\Lambda(1405)$
in the $\bar KN$ channel. We find that the 
$K^{-}d \to \Lambda(1405)n$ process
favors the production of $\Lambda(1405)$ initiated by the $\bar KN$ 
channel.  The present approach indicates that the $\Lambda(1405)$ 
resonance position is 1420 MeV rather than 1405 MeV in the $\pi\Sigma$
invariant mass spectra of $K^{-} d \to \pi \Sigma n$ reactions. 
This is consistent with an observed spectrum 
of the $K^{-} d \to \pi^{+}\Sigma^{-}n$ with 686-844 MeV/c incident $K^{-}$ 
by bubble chamber experiments done in the 70's. Our model also reproduces 
the measured $\Lambda(1405)$ production cross section.
\PACS{
      {14.20.Jn}{Hyperons} \and
      {25.80.Nv}{Kaon-induced reactions} \and
      {13.75.Jz}{Kaon-baryon interactions}  \and
      {12.39.Fe}{Chiral Lagrangians} 
     } 
\keywords{Structure of $\Lambda(1405)$ -- Kaon induced production of $\Lambda(1405)$ -- Chiral unitary model}
} 
%


\maketitle

\section{Introduction}

The structure of the $\Lambda(1405)$ resonance is an important recent
issue particularly to understand $\bar K$-nucleus interactions. 
The $\Lambda(1405)$ has been a historical example of 
a dynamically generated resonance in 
meson-baryon coupled-channels dynamics with $S=-1$~\cite{Dalitz:1967fp}.
Modern investigations based on chiral dynamics with a unitary framework 
also reproduce well the observed spectrum of the $\Lambda(1405)$ together 
with cross sections of $K^{-}p$ to various 
channels~\cite{Kaiser:1995eg,Oset:1998it,Oller:2000fj,Oset:2001cn,GarciaRecio:2002td,Hyodo:2002pk}.
Recently it was pointed out in Ref.~\cite{Hyodo:2008xr} that 
the $\Lambda(1405)$ can be regarded almost purely as 
a dynamically generated state in meson-baryon scattering,
while the description of the $N(1535)$ demands some components 
other than meson-baryon ones, such as genuine quark components. 

One of the important consequences of chiral dynamics 
is that the $\Lambda(1405)$ is described by superposition
of two resonance states~\cite{Jido:2003cb}.
One state located around 1420 MeV couples dominantly to the $\bar KN$
channel, while the other one sitting around 1390 MeV with a 130 MeV width
couples strongly to the $\pi \Sigma$ channel. Consequently,
the spectra of the $\Lambda(1405)$ depend on the channels and
the resonance position in the $\bar KN$ channel is 
1420 MeV, higher than the nominal one which is 1405 MeV. 
Therefore, it is important to observe the resonance position 
of the $\Lambda (1405)$ in the $\bar KN$ channel.

To observe the resonance position of the $\Lambda(1405)$
in the $\bar KN$ channel it is necessary to produce the $\Lambda(1405)$
by reactions initiated by $\bar KN$.
Since the $\Lambda(1405)$ resonance appears below the threshold 
of the $\bar KN$ channel, direct production of $\Lambda(1405)$ 
in the $\bar KN$ channel is kinematically forbidden. This fact 
leads us to indirect productions of $\Lambda(1405)$, such as 
$\gamma p \rightarrow \Lambda(1405) K^{*}$~\cite{Hyodo:2004vt},
$K^{-} p \to \gamma \Lambda(1405)$~\cite{nachergamma}, 
$K^{-} p \to \pi_{0} \Lambda(1405)$~\cite{prakhov,magas},
and nuclear reactions~\cite{Kishimoto:1999yj}.
Here we discuss the $K^{-}$ induced production of $\Lambda(1405)$
with a deuteron target, $K^{-} d \to \Lambda(1405) n$. In this
reaction, the final neutron takes energy out from the initial kaon 
and the $\Lambda(1405)$ is produced by the $\bar KN$ channel.  

The paper is organized as follows: In Sec.~\ref{sec:form}, we explain
our model to calculate the $K^{-} d \to \pi\Sigma n$ reactions 
and introduce the description of $\Lambda(1405)$ based on 
the chiral unitary approach. In Sec.~\ref{sec:result} we show 
our numerical results of the calculations and compare our 
results with experiments. Section \ref{sec:summary} is devoted 
to a summary of this work. 

\section{Formulation}

\label{sec:form}

In this section, we explain our approach to calculate the cross section 
of the $K^{-}d \to \pi\Sigma n$ reaction. In Sec.~\ref{sec:kine}, we
discuss the kinematics of this reaction and introduce relevant diagrams
for the $\Lambda(1405)$ production. The $T$-matrix is calculated 
in Sec.~\ref{sec:SA}. The description of the two-body meson-baryon 
scattering amplitudes and the model of $\Lambda(1405)$ in the chiral 
unitary approach are discussed in Sec.~\ref{sec:ChUA}. 

\subsection{Kinematics}
\label{sec:kine}
\begin{figure}
\centerline{\includegraphics[width=6.5cm]{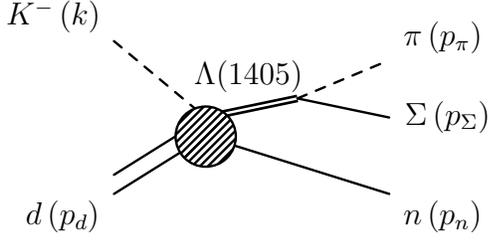}}
\caption{Kinematics of the $K^{-} d \to \pi \Sigma n$. \label{fig1}}
\end{figure}

We consider $\Lambda(1405)$ production induced by $K^{-}$ with a deuteron
target, $K^{-} d \rightarrow \Lambda(1405) n$. 
The $\Lambda(1405)$ produced in this reaction decays into $\pi \Sigma$ with $I=0$
as shown in Fig.~\ref{fig1}. The $\Lambda(1405)$ is identified by the $\pi \Sigma$ invariant mass spectra of this reaction. 
Figure~\ref{fig1} also gives the kinematical variables of
the initial and final particles. The kinematics of the three-body final state is completely
fixed by five variables, the $\pi\Sigma$ invariant mass $M_{\pi\Sigma}$, the neutron 
solid angle $\Omega_{n}$ in the c.m.\ frame and the pion solid angle $\Omega_{\pi}^{*}$
in the rest frame of $\pi$ and $\Sigma$~\cite{Amsler:2008zz}. 
Thus the differential cross section of this reaction can be written by
\begin{equation}
d \sigma = \frac{1}{ (2\pi)^{5}} 
\frac{M_{d}M_{\Sigma} M_{n}}{4k_{\rm c.m.}E_{\rm c.m.}^{2}}
   \, |{\cal T}|^{2}
   |\vec p_{\pi}^{\, *}|\,
   |\vec p_{n}|\, dM_{\pi\Sigma} d\Omega_{\pi}^{\,*} d\Omega_{n}
   \label{eq:difcross}
\end{equation}
where $\cal T$ is the $T$-matrix of this reaction, 
$E_{\rm c.m.}$ is the center of mass energy, $k_{c.m.}$ is the kaon c.m.\ momentum
and
$\vec p_{\pi}^{\, *} $ is the pion momentum in the rest frame of $\pi$ and $\Sigma$.
The pion momentum $|\vec p_{\pi}^{\, *}|$ in the $\pi \Sigma$ rest frame 
can be fixed by the invariant mass $M_{\pi\Sigma}$ as
\begin{equation}
  |\vec p_{\pi}^{\, *}| = \frac{\lambda^{1/2}(M_{\pi\Sigma}^{2}, m_{\pi}^{2}, M_{\Sigma}^{2})}{2 M_{\pi\Sigma}}
\end{equation}
with the K\"allen function $\lambda(x,y,z)=x^{2}+y^{2}+z^{2}-2xy-2yz-2zx$.

\begin{figure}
\begin{center}
\centerline{\includegraphics[width=8.5cm]{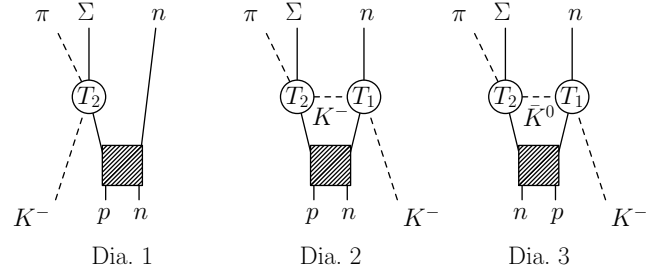}}
\caption{Diagrams for the calculation of the $K^{-}d \to \pi\Sigma n$ reaction.
$T_{1}$ and $T_{2}$ denote the scattering amplitudes for $\bar KN \to \bar KN$
and $\bar K N \to \pi \Sigma$, respectively.  \label{fig2}}
\end{center}
\end{figure}

The $\Lambda(1405)$ production is investigated by 
limiting the kinematics with the invariant mass of the final $\pi \Sigma$ state
around 1350 to 1450 MeV, in which the resonating $\pi\Sigma$ forms the 
$\Lambda(1405)$
and the resonance contribution may dominate the cross section.  
Thus, we do not consider the diagrams in which the final 
$\pi$ and $\Sigma$ are emitted from different vertices, since 
they are not correlated.
In this $\Lambda(1405)$ dominance
approximation, we have three diagrams for this reaction as shown in Fig.~\ref{fig2}.
The left diagram of Fig.~\ref{fig2} expresses  the $\Lambda(1405)$ production 
in the impulse approximation. We refer to this diagram as direct production process.  The middle and right diagrams are for two-step
processes with $\bar K$ exchange. We refer to these diagrams 
as double scattering diagrams. 

In Fig.~\ref{fig2}, $T_{1}$ and $T_{2}$ denote $s$-wave scattering amplitudes of 
$\bar K N \to \bar KN$ and $\bar K N \to \pi \Sigma$, respectively. These amplitudes 
are calculated in coupled-channels approach based on chiral dynamics, as we will 
explain later. In the amplitude $T_{2}$, the $\Lambda(1405)$ resonance is involved. Note that, in these three diagrams,  
the $\Lambda(1405)$ in the amplitude $T_{2}$ is produced 
by the $\bar KN$ channel.  
For the amplitude $T_{1}$, the energies of interest are 
1600 MeV to 1800 MeV 
for the $K^{-}$ incident momenta 600 MeV/c to 1000 MeV/c in the lab.\ frame. 
We do not consider double scattering diagrams with pion exchanges
in which $\Sigma$ and $\pi$ are emitted separately
from the $T_{1}$ and $T_{2}$ amplitudes, respectively.  
Such diagrams may give 
smooth backgrounds in the $\pi \Sigma$ invariant mass spectra. 
We do not consider the $\Sigma(1385)$ resonance in the $T_{2}$ amplitude,
since the branching rate of $\Sigma(1385)$ to $\pi\Sigma$ is only 12\%. 

For the energetic incident $K^{-}$ with several hundreds MeV/c momentum
in the lab. frame, the contribution of diagram 1 (direct production) is 
expected to be very small, since the $\Lambda(1405)$ is produced 
below the $\bar KN$ threshold by the energetic $K^{-}$ and a far off-shell 
nucleon and the deuteron wavefunction has tiny component of such a 
far off-shell nucleon. In contrast to the direct production, in the double
scattering diagrams, the large energy of the incident $K^{-}$ is carried 
away by the final neutron and the exchanged kaon can have a suitable
energy to create the $\Lambda(1405)$ colliding with the other nucleon
in the deuteron. 

\subsection{Scattering amplitude}
\label{sec:SA}
Let  us calculate the $T$-matrix for the $K^{-}d \rightarrow \pi \Sigma n$ reaction.
The $T$-matrix for the diagram 1 given in Fig.~\ref{fig2} can be calculated 
in the impulse approximation in which the incident $K^{-}$ and the proton in 
the deuteron transform into $\pi \Sigma$ and the neutron behaves 
as a spectator of the reaction. 

Letting the wavefunctions of the incident kaon and the particles in the final state be given by plane waves and writing the wavefunctions of the nucleons in the deuteron as $\varphi_{i}$ $(i=1,2)$, we obtain the connected part of the $S$-matrix:
\begin{eqnarray}
  S & = &
  \int d^{4} x_{1} \,
   (-i) T_{K^{-}p \rightarrow \pi \Sigma} \,
  \norm{\Sigma}
  e^{ip_{\Sigma} \cdot x_{1}}
  \norm{\pi}
  e^{i p_{\pi}\cdot x_{1}} \, 
   \nonumber \\ && \times \, 
   \norm{1}
   \varphi_{1}( \vec x_{1}) e^{-i p_{1}^{0} x_{1}^{0}} \,
   \norm{K^{-}}
   e^{-i k \cdot x_{1}}
   \nonumber \\ && \times
   \int d^{3} x_{2} \, 
   \norm{n} e^{-i \vec p_{n}\cdot \vec x_{2}} \, 
   \norm{2} \varphi_{2}(\vec x_{2}) 
\end{eqnarray}
where $T_{K^{-}p\rightarrow \pi \Sigma}$ is the $T$-matrix for the $K^{-}p \rightarrow \pi \Sigma$, and $p_{1}$ and $p_{2}$ are the momenta of the proton and neutron inside the deuteron, respectively. The normalization factors of the wavefunctions are given by
$
\norm{i} =   \sqrt{{M_{i}}/{E_{i}}} 
$
for baryons and 
$
\norm{a} =   {1}/{\sqrt{ 2 \omega_{a} }} 
$
for mesons. The plane waves are normalized inside a box with a unit volume. 

The integration of the time component $x_{1}^{0}$ 
give the delta function for energy conservation: 
\begin{eqnarray}
   \int dx_{1}^{0} e^{-i(k^{0}+p_{1}^{0}-p_{\Sigma}^{0}-p_{\pi}^{0}) x_{1}^{0}}
   = 2\pi \delta(k^{0}+p_{1}^{0}-p_{\Sigma}^{0}-p_{\pi}^{0}) .
\end{eqnarray}
In the impulse approximation, the energy of the spectator neutron does not 
change, $p_{2}^{0} = p_{n}^{0}$. Taking the deuteron energy as
the sum of the energies of the nucleons, $p_{d}^{0} = p_{1}^{0} + p_{2}^{0}$, 
we obtain the total energy conservation $k^{0}+p_{d}^{0}-p_{\Sigma}^{0}-p_{\pi}^{0}-p_{n}^{0}=0$. Here, in general, $p_{n}^{0} \neq p_{2}^{0}$. Thus one must assume in the impulse approximation that the initial energy in the deuteron is not equally distributed between the two nucleons. 

To perform the spacial integrals, we introduce 
the relative coordinate for $\vec x_{1}$ and $\vec x_{2}$ as
$\vec R = (\vec x_{1} + \vec x_{2})/2$ and $ \vec r = \vec x_{1} - \vec x_{2}$.
We also introduce the deuteron wavefunction for the relative motion $\varphi(r)$
and assume that the center of mass motion of the deuteron is described as 
the plane wave. Namely we replace the nucleon wavefunctions, $\varphi_{1}(\vec x_{1})$ and $\varphi_{2}(\vec x_{2})$, as follows:
\begin{equation}
   \norm{1}\norm{2} \varphi_{1}(\vec x_{1}) \varphi_{2}(\vec x_{2})
   \rightarrow \norm{d} e^{i\vec p_{d} \cdot \vec R} \varphi(\vec r)
   \label{eq:Nuc2Deu}
\end{equation}
where the deuteron wavefunction is normalized as
\begin{equation}
   \int d^{3} r | \varphi(\vec r)|^{2} = 1.
\end{equation}
We neglect $d$-wave component of the deuteron wavefunction.
The $s$-wave wavefunction $\varphi(r)$ 
in the rest frame of the deuteron
is parametrized by analytic functions~\cite{Lacombe:1981eg} as
\begin{equation}
   \varphi(r) = \sum_{j=1}^{11} \frac{C_{j}}{r} \exp(-m_{j} r)  .
\end{equation}
We use the parametrization for $C_{j}$ and $m_{j}$ given in Ref.~\cite{Machleidt:2000ge}.
The integration in terms of $\vec R$ gives  the delta function for total momentum 
conservation, while
the integration of $\vec r$ gives the Fourier transformation 
of the deuteron wavefunction:
\begin{equation}
   \int d^{3} r \varphi(r)\, e^{i(\vec k + \vec p_{n} - \vec p_{\Sigma} - \vec p_{\pi})\cdot \frac{\vec r}{2}}  = \tilde \varphi(\vec p_{n} - \frac{\vec p_{d}}{2})
\end{equation}
where we have used momentum conservation. 

Finally we obtain the $S$-matrix for the diagram 1 as
\begin{eqnarray}
   S &=& -i T_{K^{-}p \rightarrow \pi \Sigma} \, \varphi(\vec p_{n} - \frac{\vec p_{d}}{2}) \,  \norm{d} \norm{K}
   \norm{\Sigma} \norm{\pi} \norm{n} 
    \nonumber \\ && \times
    (2\pi)^{4} \delta^{4}(p_{d}+k - p_{\Sigma}-p_{\pi}-p_{n}) .
\end{eqnarray}
Since the $T$-matrix is given by $S=1 - i  (2\pi)^{4} \delta^{4}(p_{d}+k - p_{\Sigma}-p_{\pi}-p_{n}) \, (\prod_{i}\norm{i})\, {\cal T}$,
we obtain the $T$-matrix for the diagram 1 as
\begin{equation}
  {\cal T}_{1} = T_{K^{-}p \rightarrow \pi \Sigma}(M_{\pi\Sigma}) \, \varphi(\vec p_{n} - \frac{\vec p_{d}}{2}). \label{eq:T1}
\end{equation}

Next let us consider the double scattering diagrams shown as diagrams 2 and 3 in Fig.~\ref{fig2}. In the same way as the calculation of diagram 1, the connected $S$-matrix for diagram 2 is obtained as
\begin{eqnarray}
\lefteqn{ S = } && \nonumber \\ &&
   (\prod_{i} \norm{i}) 
   \int d^{4} x_{1} \int d^{4} x_{2}   \,
   (i) \int \frac{d^{4}q}{(2\pi)^{4}} \frac{e^{-iq\cdot(x_{1}-x_{2})}}{q^{2} - m_{K}^{2} + i \epsilon}
   \nonumber \\ && \times
   e^{ip_{\Sigma}\cdot x_{1}} e^{ip_{\pi}\cdot x_{1}} e^{ip_{n}\cdot x_{2}} e^{-ik\cdot x_{2}} e^{-ip_{1}^{0}x_{1}^{0}} \varphi_{1}(\vec x_{1}) 
   \nonumber \\ && \times
   e^{-ip_{2}^{0}x_{2}^{0}}\varphi_{2}(\vec x_{2})
   (-i)T_{K^{-}n \rightarrow K^{-}n}
   \, (-i) T_{K^{-}p \rightarrow \pi \Sigma}.
\end{eqnarray}
The integrations with respect to $x_{1}^{0}$ and $x_{2}^{0}$ give energy 
conservation $p_{1}^{0}+q^{0}=p_{\Sigma}^{0}+p_{\pi}^{0}$ and $p_{2}^{0}+k^{0}=q^{0}+p_{n}^{0}$, respectively. 
Integrating with respect to $q_{0}$
\begin{eqnarray}
\lefteqn{
\int\frac{dq^{0}}{2\pi} 2\pi \delta(p_{1}^{0}+q^{0}-p_{\Sigma}^{0}-p_{\pi}^{0}) 2\pi \delta(p_{2}^{0}+k^{0}-q^{0}-p_{n}^{0})
} \nonumber &&\\
&&= 2\pi \delta(k^{0}+p_{d}^{0}-p_{\Sigma}^{0}-p_{\pi}^{0}-p_{n}^{0})
\ \ \ \ \ \ \ \ \ \ \  \ \ \ \ \ \ \ \ \ \ \ \ \ 
\end{eqnarray}
we obtain total energy conservation and $q^{0}$ has been fixed as
$q^{0}= k^{0}+p_{2}^{0}-p_{n}^{0}=p_{\Sigma}^{0}+p_{\pi}^{0}-p_{1}^{0}$.
Here we have used again $p_{1}^{0}+p_{2}^{0}=p_{d}^{0}$.
Introducing again the relative coordinate for $\vec x_{1}$ and $\vec x_{2}$ and 
the deuteron wavefunction as Eq.~(\ref{eq:Nuc2Deu}),
we find that the integration with respect to $\vec R$ gives total momentum conservation and that the integration of $\vec r$ provides the Fourier transformation 
of the deuteron wavefunction:
\begin{equation}
   \int d^{3}r \, \varphi(r)\, e^{i(-\vec k + \vec p_{n} - \vec p_{\Sigma} - \vec p_{\pi}+2\vec q)\cdot \frac{\vec r}{2}}  = \tilde \varphi(\vec q+\vec p_{n}-\vec k - \frac{\vec p_{d}}{2})  \nonumber
\end{equation}
with momentum conservation $\vec p_{\Sigma} + \vec p_{\pi} = \vec p_{d} + \vec k - \vec p_{n}$. 
Finally we obtain the $T$-matrix for diagram 2 as
\begin{eqnarray}
  {\cal T}_{2}& =&  T_{K^{-}p \rightarrow \pi \Sigma}(M_{\pi\Sigma}) 
 \int \frac{d^{3}q}{(2\pi)^{3}} \frac{\tilde \varphi (\vec q+\vec p_{n}-\vec k - \frac{\vec p_{d}}{2})}{q^{2}-m_{K}^{2} + i\epsilon}
   \nonumber \\ && \times
 T_{K^{-}n \rightarrow K^{-}n}(W_{1}) \ . \label{eq:T2}
\end{eqnarray}
where $W_{1}$ denotes the invariant mass of the initial kaon and the neutron inside
the deuteron, $W_{1} \equiv \sqrt {(k+p_{2})^{2}}$. 
{In principle, $W_{1}$ depends on the integral momentum 
$\vec q$ as $W_{1}=\sqrt{(k^{0}+p_{2}^{0})^{2}-(\vec q + \vec p_{n})^{2}}$, where we have used the momentum conservation $\vec k + \vec p_{2} = \vec q + \vec p_{n}$ at the vertex. }

The variables 
{$q^{0}$ and $W_{1}$ can be} 
fixed by kinematics but depend on energies of the nucleons in the deuteron. 
To evaluate these values, let us take the deuteron rest frame, that is,
the laboratory frame. The deuteron is a loosely bound system of proton and neutron
with about 2 MeV binding energy. Neglecting the binding energy for the neutron, 
we evaluate
\begin{equation}
  q^{0} = M_{N} + k^{0} - p_{n}^{0}.  \label{eq:q0}
\end{equation}
{For the determination of $W_{1}$, we recall that the}
wavefunction of the nucleons inside the deuteron in momentum space 
has the largest component when the nucleon is almost at rest 
in the rest frame of the deuteron. 
{Namely, in Eq.~\eqref{eq:T2}, the function $\tilde \varphi$ has the maximum  
at $\vec q = \vec k - \vec p_{n}$ in the lab.\ frame.}
This fact allows us to determine $W_{1}$ as
\begin{eqnarray}
   W_{1}  = \sqrt{(M_{N}+k^{0})^{2} - \vec k^{\, 2}}.
   \label{eq:W}
\end{eqnarray}
Note that, under this assumption, the energy $W_{1}$ for the $K^{-}n \to K^{-}n$ amplitude is completely fixed by kinematics without depending on the momenta of the final state. Thus,  the value of the $K^{-}n \to K^{-}n$ scattering amplitude
contributes only to the absolute value of the cross section and does not provide
any structure in the invariant mass spectra. 
{We can estimate the deviation of $W_{1}$ from this approximation. The deuteron wavefunction~$\varphi$ has half of the maximum value at $|\vec q+\vec p_{n}-\vec k|=45$~MeV in Eq.~\eqref{eq:T2}. With this momentum, at most, $W$ can be change $\pm 20$ MeV for the 800 MeV/c incident $K^{-}$ momentum. We will discuss this approximation more in Sec.~\ref{sec:approx}.}

In the same way, we can calculate the $T$-matrix for diagram 3:
\begin{eqnarray}
  {\cal T}_{3}& =& - T_{\bar K^{0}n \rightarrow \pi \Sigma}(M_{\pi\Sigma}) 
  \int \frac{d^{3}q}{(2\pi)^{3}} \frac{\tilde \varphi (\vec q+\vec p_{n}-\vec k - \frac{\vec p_{d}}{2})}{q^{2}-m_{K}^{2} + i\epsilon}
  \nonumber \\ && \times
  T_{K^{-}p \rightarrow \bar K^{0}n}(W_{1})  \ . \label{eq:T3}
\end{eqnarray}
The negative sign in the right had side comes from the isospin configuration of the deuteron. 
{Again $W_{1}$ is fixed by Eq.~\eqref{eq:W}.}

Finally, the total $T$-matrix is given by summing up these three amplitudes:
\begin{equation}
   {\cal T} =  {\cal T}_{1} + {\cal T}_{2} + {\cal T}_{3}
\end{equation}
where ${\cal T}_{i}$ ($i=1,2,3$) are given in Eqs.~(\ref{eq:T1}), (\ref{eq:T2}) and (\ref{eq:T3}), respectively. 
Note again that, in these amplitudes ${\cal T}_{i}$ ($i=1,2,3$), 
the $\Lambda(1405)$ is involved in the $\bar KN$-initiated amplitudes
$T_{\bar K N \rightarrow \pi \Sigma}$. 

We do not consider further multiple scattering since we are dealing 
with a deuteron break up process where the final baryons go apart 
from each other and are weakly correlated. Also the initial kaon 
energy is not small. This situation is very different from the coherent 
process of $\bar K d$ elastic scattering at threshold appearing in the 
evaluation of the $\bar K d$ scattering length. In this case the slow 
kaon sticks around the nucleons of the deuteron and multiple 
scattering with the two nucleons is needed in the evaluation 
of the scattering length \cite{Toker:1981zh,Torres:1986mr,Bahaoui:1990da,Kamalov:2000iy}. 
The situation bears much resemblance to the case of elastic 
low energy pion nucleus scattering, which requires the full 
solution of the Klein-Gordon equation accounting for multiple 
scattering with the nucleons~\cite{Nieves:1991ye}, and the 
quasielastic breakup processes which proceeds incoherently with the 
contribution of the first process kinemallically allowed~\cite{Salcedo:1987md}.

\subsection{Description of the $\bar KN$ scattering amplitudes and model for the $\Lambda(1405)$}
\label{sec:ChUA}
For the description of the $\Lambda(1405)$, we use the chiral unitary approach, 
in which the $\Lambda(1405)$ is dynamically generated in coupled-channels of 
meson-baryon scattering with strangeness $S=-1$ and charge $Q=0$, namely 
$K^{-}p$, $\bar K^{0}n$, $\pi^{0} \Lambda$, $\pi^{0}\Sigma^{0}$, $\eta\Lambda$, 
$\eta\Sigma^{0}$, $\pi^{+}\Sigma^{-}$, $\pi^{-}\Sigma^{+}$, 
$K^{+}\Xi^{-}$ and $K^{0} \Xi^{0}$. The scattering amplitudes for the meson-baryon 
channels  are
obtained by solving the Bethe-Salpeter equation with the $s$-wave
interaction kernels $V$ given by chiral Lagrangian. The Bethe-Salpeter equation
turns out to be an algebraic equation of the meson-baryon coupled channels if we consider elastic unitarity in the $N/D$
method~\cite{Oller:2000fj}:
\begin{equation}
   T_{ij}(W) = V_{ij}(W) + V_{ik}(W) G_{k}(W) T_{kj}(W) . \label{eq:BSeq}
\end{equation}
where $i,j$ are the channel indices given in Ref.~\cite{Oset:1998it} and $W$ denotes the center of mass energy. 

In Eq.~(\ref{eq:BSeq}),  $V$ is given by the leading order of the chiral Lagrangian as
\begin{eqnarray}
  \lefteqn{ V_{ij}(W) =} &&  \nonumber \\
  && - \frac{C_{ij}}{4f^{2}} (2W - M_{i} - M_{j})\left(\frac{M_{i}+E_{i}}{2M_{i}}\right)^{1/2} \left(\frac{M_{j}+E_{j}}{2M_{j}}\right)^{1/2}  \label{eq:WTterm}
\end{eqnarray}
with the baryon masses $M_{i}$ and energy $E_{i}$. The coefficient $C_{ij}$ in
Eq.~(\ref{eq:WTterm}) is the channel coupling which is fixed by the flavor SU(3) group structure, and $f$ is the decay constant of the meson in the chiral field $U=\exp[i\sqrt 2 \Phi /f ]$. Here we use an averaged value of
$f=1.123 f_{\pi}$ with $f_{\pi}=93$ MeV
as done in Ref.~\cite{Oset:2001cn} to deal with the $\bar K N$ interaction. 

The diagonal matrix $G_{k}(W)$ in Eq.~(\ref{eq:BSeq}) is a meson-baryon loop function and is evaluated with dimensional regularization:
\begin{eqnarray}
\lefteqn{
   G_{k}(W) =
    i \int \frac{d^{4}q}{(2\pi)^{4}} \frac{2M_{k}}{(P-q)^{2}-M_{k}^{2}+i\epsilon} \frac{1}{q^{2}-m_{k}^{2}+i\epsilon} } && \nonumber \\
   &= &\frac{2M_{k}}{16\pi^{2}} \left\{ a_{k}(\mu) + \ln \frac{M_{k}^{2}}{\mu^{2}}
   + \frac{m_{k}^{2}-M_{k}^{2}+W^{2}}{2W^{2}} \ln \frac{m_{k}^{2}}{M_{k}^{2}}
   \nonumber  \right. \\ &&
   + \frac{\bar q_{k}}{W} \left[ \ln(W^{2}-(M_{k}^{2}-m_{k}^{2})+2\bar q_{k} W)
   \nonumber  \right. \\ &&
   +\ln(W^{2}+(M_{k}^{2}-m_{k}^{2})+2\bar q_{k} W)
   \nonumber \\ &&
  - \ln(-W^{2}+(M_{k}^{2}-m_{k}^{2})+2\bar q_{k} W)
   \nonumber \\ && \left.\left.
   - \ln(-W^{2}-(M_{k}^{2}-m_{k}^{2})+2\bar q_{k} W) \right] \right\}
\end{eqnarray}
where  $m_{i}$ is the meson mass,  $\bar q_{i}$ is the center of mass momentum
and $a_{i}(\mu)$ is the subtraction constant with a renormalization scale $\mu$. 
The subtraction constants are only the free parameters in this model and are determined so as to reproduce the threshold properties of $K^{-}p$ to several channels~\cite{Kaiser:1995eg}, which were obtained by $K^{-}$ absorptions 
of Kaonic hydrogen~\cite{threshold}. In a recent work~\cite{Hyodo:2008xr},
it is found that the values of the subtraction constant are very important to determine the nature of the dynamically generated resonances in the chiral unitary approach. Here we take a standard parameter set given in Ref.~\cite{Oset:2001cn}:
\begin{eqnarray}
\begin{array}{ccc}
  a_{\bar KN} = -1.84,\ \  & a_{\pi\Sigma}=-2.00,\ \  & a_{\pi \Lambda} = -1.83, \ \ \\
  a_{\eta \Lambda} = -2.25, \ \ & a_{\eta\Sigma}=-2.38, \ \ & a_{K\Xi} = - 2.67, \ \ 
\end{array}  \label{eq:subtpara}
\end{eqnarray}
with $\mu = 630$ MeV.
The scattering amplitudes calculated with this parameter set reproduce well
the $\pi\Sigma$ invariant mass spectrum and the total cross sections of $K^{-}p$ to several channels~\cite{Oset:2001cn},
and the $\Lambda(1405)$ is found to be almost a purely dynamical state of meson and baryon~\cite{Hyodo:2008xr}. 

We use the scattering amplitudes obtained in the chiral unitary approach described above for the $\bar K N \rightarrow \bar K N$ and $\bar K N \rightarrow \pi \Sigma$ amplitudes in Eqs.~(\ref{eq:T1}), (\ref{eq:T2}) and (\ref{eq:T3}).
After fixing the subtraction parameters $a_{i}(\mu)$ in the two-body scattering,
we have no adjustable parameters for the calculation of the $K^{-}p \to \pi\Sigma n$ reaction in the present approach. 

\begin{figure}
\centerline{\includegraphics[width=7.5cm]{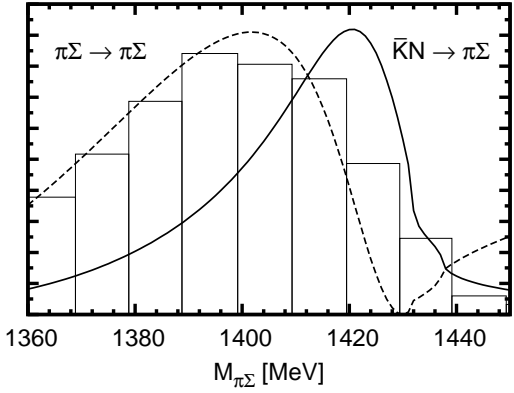}}
\caption{$\pi \Sigma$ invariant mass spectra of two-body scatterings of  
$\bar KN \rightarrow \pi\Sigma$ (solid line) and 
$\pi\Sigma \rightarrow \pi \Sigma$ (dashed line)
with $I=0$ in arbitrary units.
The histogram denotes an experimental data in Ref.~\cite{Hemingway:1984pz}.
\label{fig:MSI0}}
\end{figure}

The scattering amplitudes obtained in the chiral unitary approach have notable features. The $\Lambda(1405)$ is successfully reproduced by meson-baryon dynamics, but the resonance position depends on the channels~\cite{Jido:2003cb}. 
In Fig.~\ref{fig:MSI0}, we plot the $\pi\Sigma$ invariant mass distributions 
for $\bar KN \rightarrow \pi\Sigma$ and $\pi\Sigma \rightarrow \pi\Sigma$ with $I=0$ defined by
\begin{equation}
   \frac{d\sigma_{MB}}{dM_{\pi\Sigma}} = A |T|^{2} q_{\rm c.m} \label{eq:IMdist}
\end{equation}
where $A$ is  a constant, $q_{\rm c.m.}$ is the CM momentum of the final $\pi\Sigma$ state and $T$ is the two-body meson-baryon scattering amplitude. 
We also show in Fig.~\ref{fig:MSI0}, as the histogram, the experimental data of the $\pi^{-}\Sigma^{+}$ invariant mass spectrum obtained in $K^{-}p \rightarrow \Sigma^{+}\pi^{-}\pi^{+}\pi^{-}$ at 4.2 GeV/c for a $K^{-}$ beam with the restriction of the $\Sigma^{+}\pi^{-}\pi^{+}$ invariant mass being between 1.6 to 1.72 GeV at which the $\Sigma(1660)$ resonance (acting as a doorway of the reaction) is located~\cite{Hemingway:1984pz}.

As seen in Fig.~\ref{fig:MSI0}, the line shapes of these two channel are significantly different and the peak position in the $\pi\Sigma$ initiated scattering is 20 MeV lower than in $\bar KN$ scattering. This is because $\pi\Sigma$ is so strongly correlated as to produce the lower mass resonance pole in these energies\footnote{The importance of the strong $\pi\Sigma$ correlation in $\bar KN$ subthreshold scattering with $S=-1$ was recently pointed out in Ref.~\cite{{Hyodo:2007jq}}.}. As already discussed above, in the present process $K^{-} d \rightarrow \Lambda(1405) n$, the $\Lambda(1405)$ is driven by the $\bar KN$ channel. Thus, it is expected that the $\Lambda(1405)$ spectrum shape has a peak around 1420 MeV instead of 1405 MeV. 

\begin{figure}
\centerline{\includegraphics[width=7.5cm]{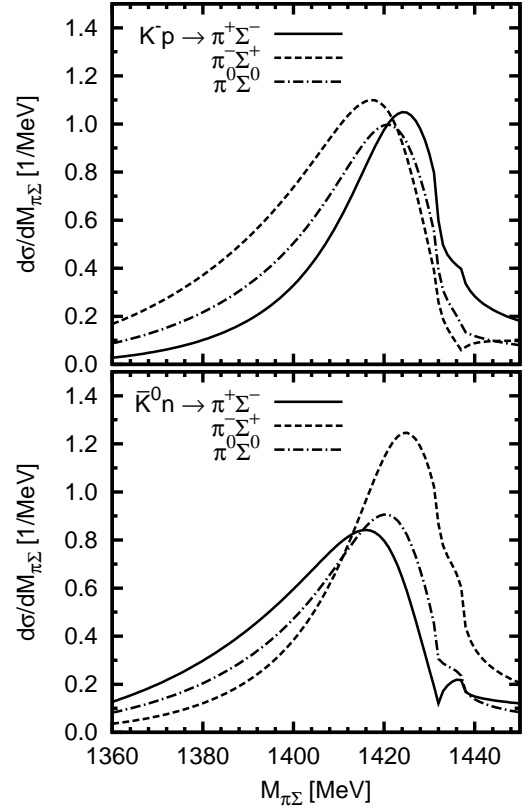}}
\caption{$\pi \Sigma$ invariant mass spectra of two-body scatterings of  
$K^{-}p \rightarrow \pi\Sigma$ (upper panel) and 
$\bar K^{0} n \rightarrow \pi \Sigma$ (lower panel)
for different $\pi\Sigma$ charged states. We take $A=1$ in Eq.~(\ref{eq:IMdist}).
\label{fig:MSKN}}
\end{figure}

We also remark that the $\Lambda(1405)$ line shapes are moderately different 
in different charged states of the $\pi$ and $\Sigma$. This is simply due to the interference of $I=0$ and $I=1$ components in the scattering amplitudes~\cite{nacherphoto}. For later convenience we show the mass spectra (\ref{eq:IMdist}) for $K^{-}p$ and $\bar K^{0}n$ to $\pi^{\pm}\Sigma^{\mp}$ and $\pi^{0}\Sigma^{0}$ in Fig.~\ref{fig:MSKN}.

\begin{figure}
\centerline{\includegraphics[width=7.5cm]{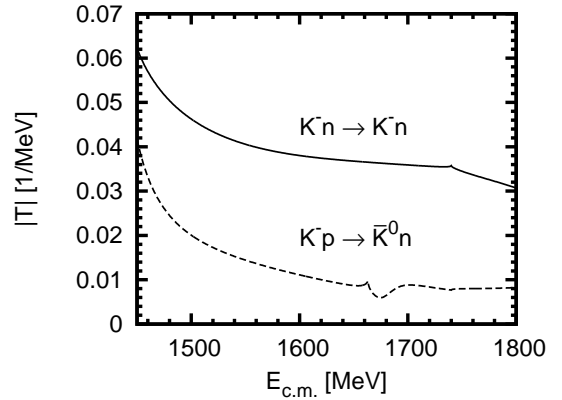}}
\caption{Modules of $\bar KN$ scattering amplitude obtained in the chiral unitary approach.
\label{fig:ScattAmp}}
\end{figure}

For the amplitude $T_{1}$ in Fig.~\ref{fig2}, we plot, in Fig.~\ref{fig:ScattAmp}, the modules of the $\bar KN$ scattering amplitudes in the relevant energy region to the present calculation. The $K^{-}n \to K^{-}n$ amplitude appearing in diagram 2 is about a factor 4 larger than $K^{-}p \to \bar K^{0}n$ in diagram 3
for the energy range from $E_{\rm c.m.}=1600$ to 1800 MeV.
Thus, we expect that diagram 2 gives a larger contribution than diagram 3
for the incident $K^{-}$ momenta from 600 to 1000 MeV/c.
The dip structure around 1670 MeV in $K^{-}p \to \bar K^{0} n$ is due to the presence of the $\Lambda(1670)$ resonance with $I=0$, while $K^{-}n \to K^{-}n$ has purely $I=1$ and there is no structure for the $\Lambda(1670)$.

\section{Results}

\label{sec:result}

In this section, we show the numerical results of the calculations for
the $K^{-} d \rightarrow \pi \Sigma n$ reaction. 
The $\pi\Sigma$ invariant mass is calculated by 
integrating the differential cross section (\ref{eq:difcross}) with respect 
to the angles of the final pion and neutron: 
\begin{equation}
\frac{d \sigma}{dM_{\pi\Sigma}} = 
\frac{M_{d}M_{\Sigma} M_{n}}{ (2\pi)^{3} \, 2k_{\rm c.m.}E_{\rm c.m.}^{2}}
   \, \int |{\cal T}|^{2}
   |\vec p_{\pi}^{\, *}|\,
   |\vec p_{n}|\, d\cos\theta  \label{eq:IMspec}
\end{equation}
%
where $\theta$ is the scattering angle of the neutron in the c.m. frame.
The integrals in terms of the pion solid angle and neutron azimuth can be
performed, since the amplitude is not dependent on these angles.
The $T$-matrix $\cal T$ is evaluated as a sum of 
the amplitudes given in Eqs.~(\ref{eq:T1}), (\ref{eq:T2}) and (\ref{eq:T3}).
These amplitudes contain two-body meson-baryon scattering amplitudes of
$\bar K N \rightarrow \bar K N$ and $\bar K N \rightarrow \pi \Sigma$.
These amplitudes are calculated in the chiral unitary approach described 
in Sec.~\ref{sec:ChUA}. 
The $\bar KN \to \bar KN$ scattering amplitudes are functions of the c.m. energy of the incident kaon and one of the nucleons in the deuteron, which can be determined when the initial kaon momentum is fixed, as discussed
in Eq.~(\ref{eq:W}), while the $\bar KN \to \pi\Sigma$ amplitudes are
functions of the $\pi\Sigma$ invariant mass $M_{\pi\Sigma}$ and 
provide the resonance shape for the $\Lambda(1405)$.

\subsection{Comparison with the experimental data of $K^{-} d \to \pi^{+} \Sigma^{-} n$}

First of all, we compare our theoretical calculation of the $K^{-}d \to \pi^{+}\Sigma^{-}n$ reaction with the experimental data reported in Ref.~\cite{Braun:1977wd}.
The experiment was performed with $K^{-}$ having momenta between 
686 and 844 MeV/c and the particles were detected by bubble chamber. 

\begin{figure}
\centerline{\includegraphics[width=7.5cm]{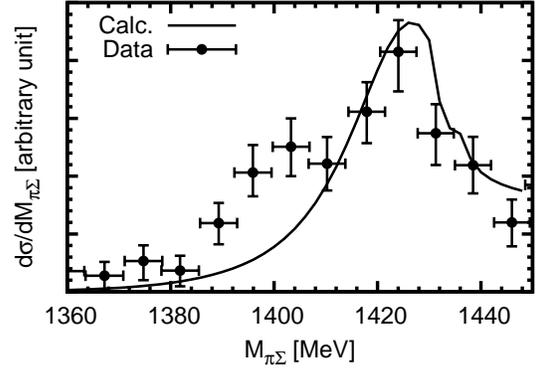}}
\caption{$\pi \Sigma$ invariant mass spectra of $K^{-}d \rightarrow \pi^{+}\Sigma^{-}n$ in arbitrary units
at 800 MeV/c incident $K^{-}$ momentum.
The solid line denotes the present calculation.
The data are taken from the bubble chamber experiment at $K^{-}$ momenta between 686
and 844 MeV/c given in Ref.~\cite{Braun:1977wd}.
\label{fig:IMspec}}
\end{figure}

\subsubsection{$\pi^{+}\Sigma^{-}$ invariant mass spectrum}
We show, in Fig.~\ref{fig:IMspec},
the $\pi^{+}\Sigma^{-}$ invariant-mass spectrum 
in arbitrary units at 800 MeV/c incident $K^{-}$ momentum and compare 
our theoretical calculation with the experimental data.  
The data are taken from  the bubble chamber experiment at $K^{-}$ momenta between 
686 and 844 MeV/c~\cite{Braun:1977wd}. 
In the analysis of this experiment, the resonance contribution was determined by fitting 
a relativistic Breit-Wigner distributions and a smooth background 
parametrized as a sum of Legendre polynomials to the data. 
We show,  in Fig.~\ref{fig:IMspec}, the resonance (foreground) contributions for 
the experimental data which are obtained by  
subtracting the background contributions estimated with the Legendre 
polynomials in Ref.~\cite{Braun:1977wd} from the actual data points given in 
the paper.
The solid line denotes our theoretical calculation with the chiral unitary approach. 

The spectrum shape obtained in this calculation agrees with that of the experimental observation.
Especially it is very interesting to see that the peak position, which comes from the $\Lambda(1405)$
production, appears around $M_{\pi\Sigma}=1420$ MeV instead of 1405 MeV
announced nominally by the Particle Data Group. 
This is one of the strongest evidences that the resonance position of the $\Lambda(1405)$ depends on the initial channel of meson and baryon,
and supports the double pole nature of the $\Lambda(1405)$,
in which the higher state sitting in 1420 MeV strongly couples to the $\bar KN$ channel. 

\begin{figure}
\centerline{\includegraphics[width=7.5cm]{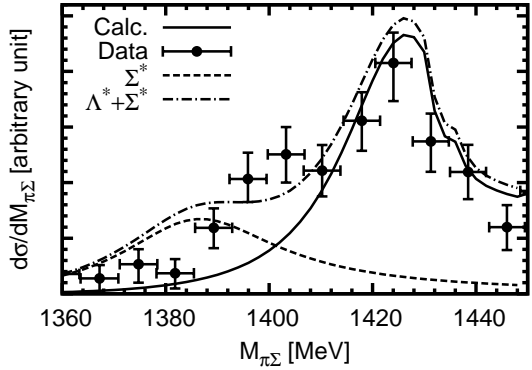}}
\caption{
A possible inference of the $\Sigma(1385)$ resonance to the 
$\Lambda(1405)$ spectrum of $K^{-}d \rightarrow \pi^{+}\Sigma^{-}n$ 
at 800 MeV/c incident $K^{-}$ momentum. 
The $\Sigma(1385)$ spectrum (dashed line) is calculated 
by the Breit-Wigner amplitude and the phase space factor 
$|\vec p_{\pi}^{\, *}|\,|\vec p_{n}|$. 
The dash-dotted lines denotes an incoherent sum of the $\Lambda(1405)$ 
and $\Sigma(1385)$ spectrum. 
The height of the $\Sigma(1385)$ spectrum is adjusted so as to reproduce
the bump structure around 1390 MeV in the observed spectrum. 
See also the caption of Fig.~\ref{fig:IMspec}.
\label{fig:IMspecSig}}
\end{figure}

The bump structure seen around $M_{\pi\Sigma}=1390$~MeV is probably 
related to the
$p$-wave contributions coming from the $\Sigma^{*}(1385)$ production, 
which we did not take into account in the present calculation. 
Let us estimate a possible 
influence of the $\Sigma(1385)$ resonance on 
the $\Lambda(1405)$ spectrum appearing around 1420~MeV. 
We calculate the $\Sigma(1385)$ spectrum in the Breit-Wigner 
formulation with the mass 1385 MeV and the width 37~MeV
including the phase space factor $|\vec p_{\pi}^{\, *}|\,|\vec p_{n}|$
as seen in Eq.~(\ref{eq:IMspec}). 
Summing up the spectra of the $\Lambda(1405)$ and $\Sigma(1385)$ 
incoherently, we find that the peak structure at 1420~MeV
is not affected by the $\Sigma(1385)$ contribution as seen in 
Fig.~\ref{fig:IMspecSig} (dash-dotted line). 
In this estimation, we have adjusted the height
of the $\Sigma(1385)$ spectrum so as to reproduce the observed 
bump structure around 1390 MeV. 
For further quantitative calculations of the $\Sigma(1385)$ spectrum,
we would have to take into account of the $\Sigma(1385)$ production 
mechanism in the present reaction and sum the amplitudes coherently,
which could be done by including the $p$-wave contributions with 
the $\Sigma(1385)$ resonance in the present formulation of 
the two-body meson-baryon scattering amplitudes 
$T_{\bar K N \to \pi \Sigma}$ following Ref.~\cite{Jido:2002zk}.
Because in angular integrated cross sections the interference of 
$s$- and $p$-waves disappears, the incoherent sum done here should 
be good and the relevant finding for the present work is that 
consideration of  the $\Sigma(1385)$ contribution does not distort 
the signal of the $\Lambda(1405)$ that we find.

Since the purpose of the paper is to show that the peak around 1420 MeV 
is a reflection of the $\Lambda(1405)$ and is narrower than for the 
nominal $\Lambda(1405)$, we have done an alternative study 
assuming the shape of Fig.~\ref{fig:IMspecSig} to be made 
by two Breit Wigner distributions that add incoherently. A fit of excellent 
quality is obtained and the structure peaking around 1420 MeV is very 
similar to what we obtain theoretically from the $\Lambda(1405)$.  
A best fit with only one Breit Winger structure, obviously does not 
reproduce the peak at lower energies in Fig.~\ref{fig:IMspecSig} and 
is of lower quality than that of the two structures. Even then, 
it is interesting to mention that the peak of this only structure is 
still around 1420  MeV.

\subsubsection{$\Lambda^{*}$ production cross section}
We also  estimate the production cross section of the $\Lambda(1405)$ in the 
$K^{-}d \rightarrow \Lambda(1405) n$ 
reaction by integrating the spectrum obtained in our calculation over
the $\pi\Sigma$ invariant mass around the resonance peak:
\begin{equation}
  \sigma_{\Lambda^{*}} = 3 \int_{M_{\rm min}}^{M_{\rm max}} d M_{\pi^{+} \Sigma^{-}} 
  \frac{d\sigma}{d M_{\pi^{+} \Sigma^{-}}} \ .
\end{equation}
The factor 3 accounts for the branching ratio of 
$\Lambda(1405) \to \pi^{+}\Sigma^{-}$.
Taking $M_{\rm min}=1400$ MeV and $M_{\rm max}=1440$ MeV read from 
the figure, we obtain the $\Lambda(1405)$ 
production cross section as 385 $\mu$b with 800 MeV/c incident $K^{-}$. 
An experimental value observed in the 
$K^{-}d \to \pi^{+}\Sigma^{-}n$ reaction is reported to be 410$\pm$100 
$\mu$b at 778 MeV/c of the incident $K^{-}$ 
momentum~\cite{Braun:1977wd}. 
The present calculation fairly agrees with the observed value. 
This implies that the $\Lambda(1405)$ production mechanism in the present
reaction is explained by the three diagrams shown in Fig.~\ref{fig2}. 

\begin{figure}
\centerline{\includegraphics[width=7.5cm]{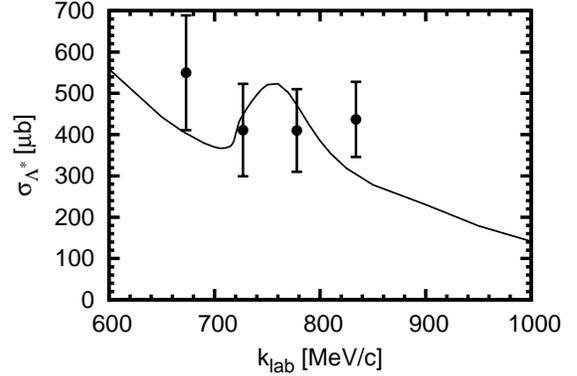}}
\caption{ Incident $K^{-}$ momentum dependence of the $\Lambda(1405)$ 
production cross section calculated with the $K^{-}d \to \pi^{+} \Sigma^{-} n$
reaction. The data are taken from Ref.~\cite{Braun:1977wd}.
\label{fig:ProdCSLambda}}
\end{figure}

In Fig.~\ref{fig:ProdCSLambda}, we show the incident momentum dependence 
of the $\Lambda(1405)$ production cross section. The cross sections are evaluated 
from the $K^{-}d \to \pi^{+} \Sigma^{-} n$ channel by integrating the invariant
mass spectra from 1400 MeV to 1440 MeV and multiplying by the isospin factor 3. 
The experimental data are taken again from Ref.~\cite{Braun:1977wd}.
Our calculation is consistent with the experimental data. 
The bump structure seen in the theoretical calculation  around $k_{\rm lab}=750$ MeV/c corresponds to the $\Lambda(1670)$ resonance production
in the $T_{1}$ amplitude of $K^{-}p \rightarrow \bar K^{0} n$
(see Fig.~\ref{fig2})~\footnote{The bump structure could be less 
pronounced with corrections to the factorization approximation 
of the $T_{K^{-} N \to \bar KN}$ amplitude
for the double scattering diagrams done in Eqs.~(\ref{eq:T2}) 
and (\ref{eq:T3}). For the details, see Sec.~\ref{sec:approx}.}. 

\subsection{Theoretical results of the $\pi\Sigma$ invariant mass spectra}

\subsubsection{Spectra of other $\pi\Sigma$ charged states}

\begin{figure}
\centerline{\includegraphics[width=7.5cm]{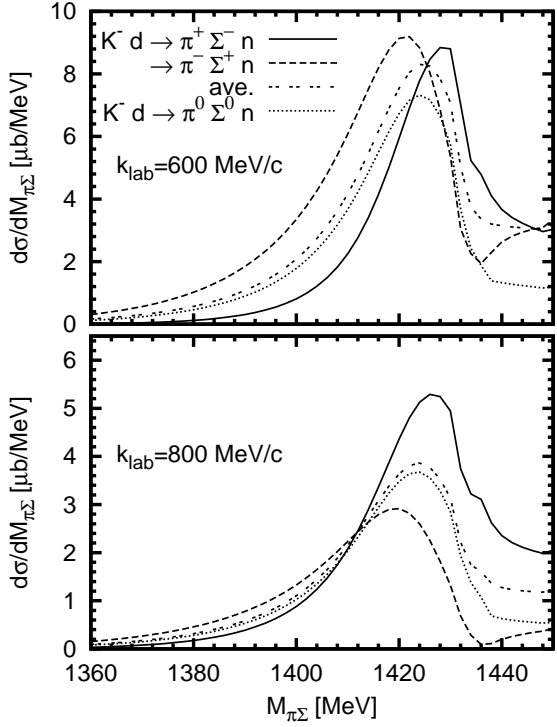}}
\caption{$\pi \Sigma$ invariant mass spectra of $K^{-}d \rightarrow \pi\Sigma n$ with 
different charged $\pi\Sigma$ states.
The solid, dashed and dotted lines denote the invariant mass spectra of $\pi^{+}\Sigma^{-}$,
$\pi^{-}\Sigma^{+}$ and $\pi^{0}\Sigma^{0}$ for the final $\pi \Sigma$ state, respectively. 
The double-dotted line shows the averaged spectrum of $\pi^{+}\Sigma^{-}$ and 
$\pi^{-}\Sigma^{+}$ of the final $\pi\Sigma$ states. 
\label{fig:IMspecPiSig}}
\end{figure}

In the previous section, we have seen that the $\Lambda(1405)$ resonance appears around 1420 MeV in the $\pi^{+}\Sigma^{-}$ invariant mass spectrum. 
To confirm that the resonance position of the $\Lambda(1405)$ in the $\bar K$ induced processes is higher than the nominal $\Lambda(1405)$,
it is certainly necessary to project out the $I=0$ contributions from the invariant mass spectra, since the shift of the resonance position could be explained 
by interference between the $I=0$ $\Lambda(1405)$ resonance and non-resonant $I=1$ $\pi\Sigma$ correlations. 

In Fig.~\ref{fig:IMspecPiSig}, we plot the $\pi\Sigma$ invariant mass spectra for the different charged $\pi\Sigma$ states with incident $K^{-}$ momenta of 600 and 800 MeV/c. We also plot the averaged spectrum of $K^{-}d \rightarrow \pi^{+}\Sigma^{-} n$ and $K^{-}d \rightarrow \pi^{-}\Sigma^{+} n$, in which interference terms of $I=0$ and $I=1$ in the
$\pi \Sigma$ correlation are cancelled out. In the spectrum of $K^{-}d \rightarrow \pi^{0}\Sigma^{0} n$,
there are not $I=1$ contributions in the $\pi \Sigma$ system. 

The peak positions are slightly dependent on the charged channels
due to the interference between the $\Lambda(1405)$ resonance
with $I=0$ and non-resonant $\pi\Sigma$ scattering with $I=1$. 
The differences are within several MeV. The peak positions are 
insensitive to the incident $K^{-}$ momenta, but the heights 
are dependent on the $K^{-}$ momenta. This is because the contribution
from each diagram depends on the incident momentum and 
interference between the diagrams shown in Fig.~\ref{fig2}
makes the heights of the spectrum dependent on the incident momentum.

\subsubsection{Contributions from each diagram}

\begin{figure}
\centerline{\includegraphics[width=9cm]{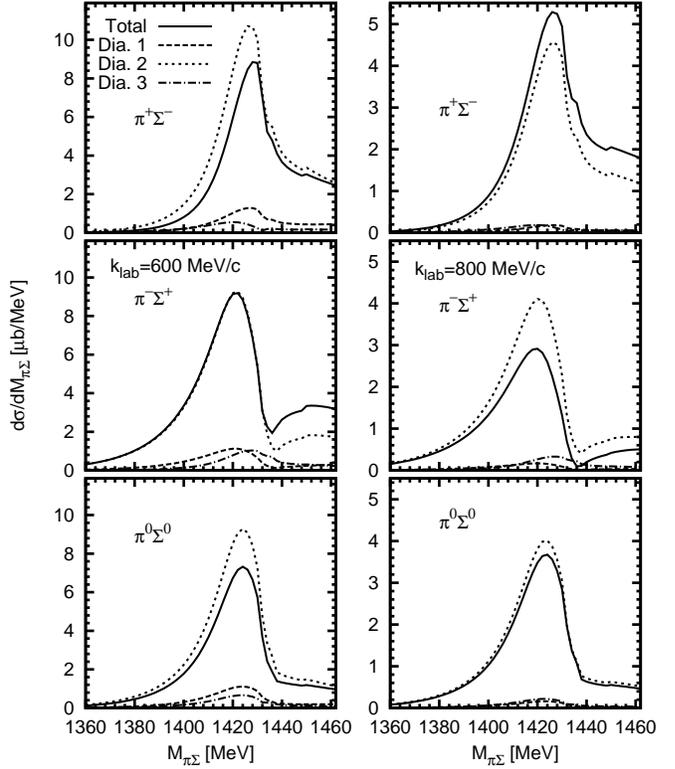}}
\caption{$\pi \Sigma$ invariant mass spectra of $K^{-}d \rightarrow \pi\Sigma n$ 
separately plotted in each diagram contribution.
The solid line denotes the total contributions of three diagrams. 
The dashed, dotted and dash-dotted lines show the calculations from diagram 1, 2 and 3
shown in Fig.~\ref{fig2}, respectively. The plots in the left and right panels 
are calculated with 600 and 800 MeV/c of $K^{-}$ incident momenta,
respectively. The charge states of $\pi \Sigma$ are shown in the plots. 
\label{fig:IMspecDia}}
\end{figure}

In Fig.~\ref{fig:IMspecDia} we show the $\pi\Sigma$ invariant mass spectra 
with separated contributions from the diagrams  shown in Fig.\ref{fig2}.
In the diagram 1, the $\Lambda(1405)$
resonance is created by the incident $K^{-}$ and the proton inside the deuteron,
and the neutron is a spectator of the $\Lambda(1405)$ production. In the
diagrams 2 and 3, two nucleons contribute the $\Lambda(1405)$ production
with one kaon exchange. 
In Fig.~\ref{fig:IMspecDia}, we find that the diagram~2 gives the dominant contribution
to the total spectrum and that the impulse approximation diagram 1 is negligible,
specially at higher incident momenta. 
The reason why the two-nucleon processes give large contributions  
can be understood as follows. In the present reaction the $\Lambda(1405)$
is produced by the $\bar K N$ channel below its threshold. Thus, some amount 
of energy should be taken out by the final neutron with energy transfer 
between the nucleons.
In the  double scattering processes (diagrams~2 and 3), 
the transferred energy is taken from the initial kaon, such that the exchanged
kaon carries less energy than on shell. 
%
%
In the impulse process of the diagram~1, the energy should be transferred by
the relative motion of the nucleons inside the deuteron. Due to the small 
binding energy of the deuteron, the deuteron wavefunction has small 
high energy components. Therefore, the contribution of the diagram~1 is
negligibly small. The reason that the diagram 3 is relatively smaller than 
diagram 2 is that the $s$-wave $K^{-} p \rightarrow \bar K^{0} n$ amplitude in the
diagram 3 is smaller than the $K^{-} n \rightarrow K^{-} n$ in the diagram 2
at the present energy region as seen in Fig.~\ref{fig:ScattAmp}. 

\subsubsection{Angular dependence of the $\Lambda(1405)$ production 
cross section}

\begin{figure}
\centerline{\includegraphics[width=7.5cm]{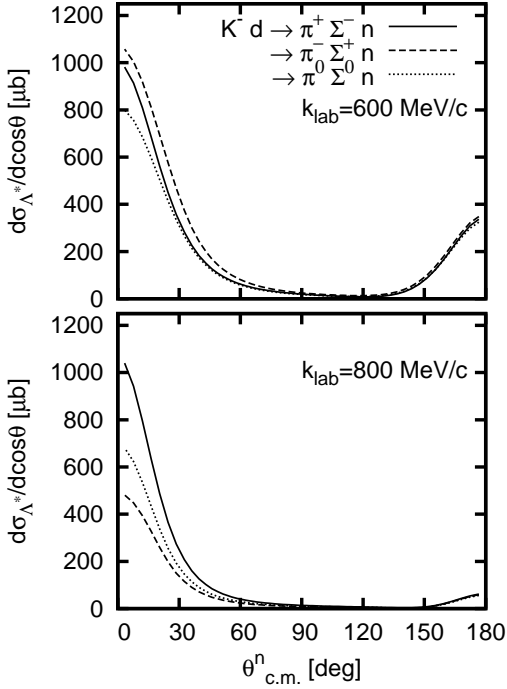}}
\caption{ Angular dependence of the $\Lambda(1405)$ production 
cross section. The horizontal axis is the angle between the incident kaon and  
the emitted neutron in center of mass frame. The incident $K^{-}$ momenta
are 600 MeV/c and 800 MeV/c in the upper and lower panels, respectively.
\label{fig:angprod}}
\end{figure}

In Fig.~\ref{fig:angprod}, we plot the angular dependence of the 
$\Lambda(1405)$ production cross section of a function of 
the angle of the incident kaon and the outgoing neutron 
in the center of mass frame. The differential cross section is 
calculated by
\begin{equation}
   \frac{d\sigma_{\Lambda^{*}}}{d\cos \theta}
   = \int_{M_{\rm min}}^{M_{\rm max}} d M_{\pi \Sigma} 
  \frac{d\sigma}{d M_{\pi \Sigma} d\cos \theta}
\end{equation}
with $M_{\rm min}=1400$ MeV and $M_{\rm max}=1440$ MeV.
Here we do not multiply by a factor 3, since we consider the production
cross section of each final $\pi\Sigma$ state. 

There are two contributions for the $\Lambda(1405)$ production. One is 
backward $\Lambda^{*}$ production ($\theta_{\rm c.m.}^{n} \simeq 0^{\circ}$)
and the other one is forward production 
($\theta_{\rm c.m.}^{n} \simeq 180^{\circ}$). 
Figure~\ref{fig:angprod} shows that large contributions to the total 
$\Lambda(1405)$ production come from the backward production.

The forward $\Lambda^{*}$ production with 
$\theta_{\rm c.m.}^{n} > 150^{\circ}$ comes from diagram~1 in Fig.~\ref{fig2}.
As already mentioned, in this diagram the neutron inside the deuteron 
can be regarded as a spectator of the $\Lambda(1405)$ production from
$K^{-}p$. Therefore the neutron 
goes to the backward angles $\theta_{\rm c.m.}^{n} > 150^{\circ}$.
On the other hand, in the double scattering processes (diagram~2 and 3),
the energetic incident kaon kicks one of the nucleons in the deuteron to forward 
directions and the exchanged kaon goes to backward angles.
The recoil momentum depends on the emitted nucleon angles and 
gets smaller with larger recoil angles.
The exchanged kaon going backward produces the $\Lambda(1405)$ together
with the other nucleon in the deuteron.
Since the momentum distribution of the nucleons in the deuteron 
is concentrated at low momenta, less than 100 MeV/c, 
a good momentum matching to create the $\Lambda(1405)$ is achieved 
by small momenta of the exchanged kaon, namely large momentum transfer 
of the incident $K^{-}$ to the emitted neutron. Therefore, in the 
double scattering processes,  the production rate of the $\Lambda(1405)$
is dominated by the backward angles.  

\subsubsection{Incident $K^{-}$ momentum dependence}

As we have already shown in Fig.~\ref{fig:ProdCSLambda}, 
the $\Lambda(1405)$ production cross section in $K^{-} d \rightarrow 
\pi^{+} \Sigma^{-} n$ decreases as the incident $K^{-}$ momentum 
increases. This is because more energetic incident $K^{-}$
produce the $\Lambda(1405)$ less efficiently due to 
worse momentum matching between the kaons and the nucleons 
inside the deuteron in both direct production (diagram~1) and 
double scattering (diagram~2 and 3) processes. 

\begin{figure}[t]
\centerline{\includegraphics[width=7.5cm]{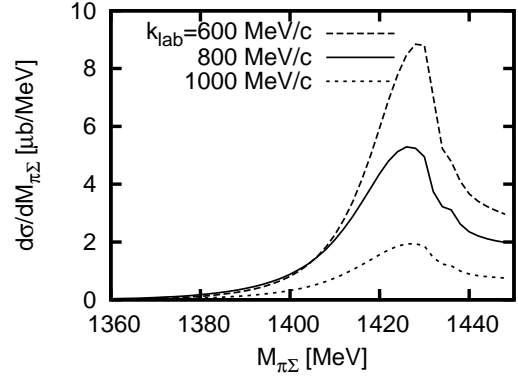}}
\caption{ $\pi \Sigma$ invariant mass spectra of 
 the $K^{-}d \to \pi^{+} \Sigma^{-} n$ reaction with 600, 800 
 and 1000 MeV/c incident $K^{-}$
 momenta. 
\label{fig:IMspecene}}
\end{figure}

In Fig.~\ref{fig:IMspecene}, we show the incident momentum dependence of 
the $\pi \Sigma$ invariant spectra of the 
$K^{-}d \to \pi^{+} \Sigma^{-} n$ reaction.
We find that the height of the resonance peak for the $\Lambda(1405)$ 
around 1420 MeV has a strong incident momentum dependence, as seen 
in the production cross section, but the spectrum shapes do not 
depend on the incident energy.

\begin{figure}
\centerline{\includegraphics[width=9cm]{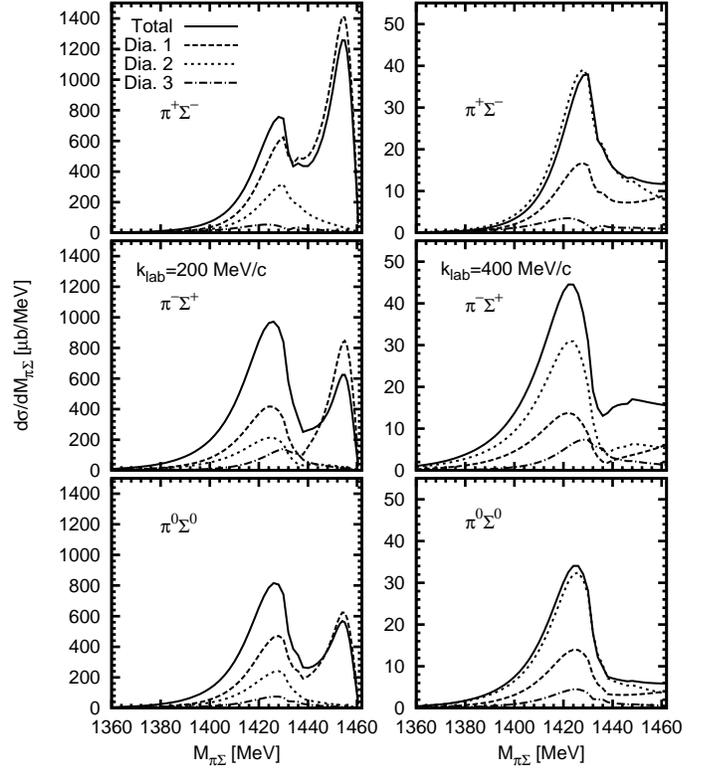}}
\caption{$\pi \Sigma$ invariant mass spectra of $K^{-}d \rightarrow \pi\Sigma n$ 
separately plotted in each diagram contribution.
The plots in the left and right panels 
are calculated with 200 and 400 MeV/c of $K^{-}$ incident momenta,
respectively. Same as described in the caption to 
Fig.~\ref{fig:IMspecDia}. \label{fig:IMspecDialow}}
\end{figure}

\begin{figure}
\centerline{\includegraphics[width=7.5cm]{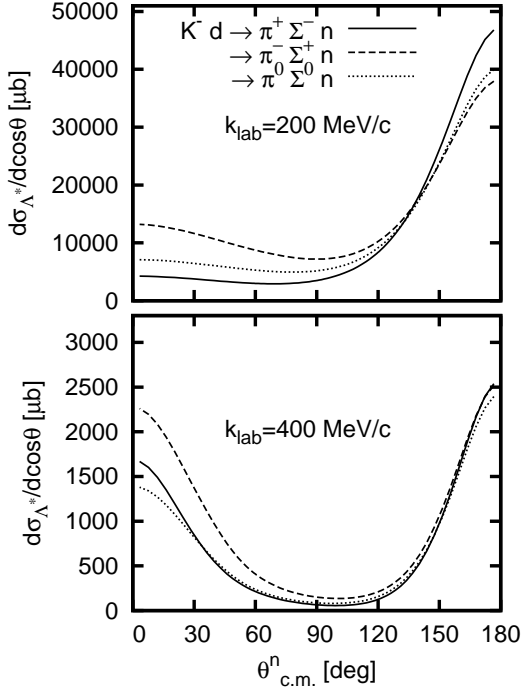}}
\caption{ Angular dependence of the $\Lambda(1405)$ production 
cross section for the incident $K^{-}$ momenta
of 200 MeV/c and 400 MeV/c in the upper and lower panels, respectively.
Same as described in the caption to Fig.~\ref{fig:angprod}.
\label{fig:angprodlow}}
\end{figure}

For lower momenta of the incident $K^{-}$, since the production 
cross sections of the $\Lambda(1405)$ are enhanced, one would expect
some advantage to observe the $\Lambda(1405)$ using low energy 
incident $K^{-}$.
This is true for not very low momenta of 
the incident $K^{-}$. For instance, we plot the $\pi\Sigma$ invariant
mass spectra at  400 MeV/c incident $K^{-}$ momentum 
in the right panel of Fig.~\ref{fig:IMspecDialow}. These figures show
that the heights of the spectra are about five times larger than 
those of the spectra obtained with 600 MeV/c of the incident $K^{-}$,
and that the spectrum shapes are very similar to each other.
Thus, using 400 MeV/c
incident $K^{-}$ is better to observe the $\Lambda(1405)$ spectra. 
It is also seen that diagram 1 and 3 give a moderate contribution
to the spectra, as seen in Fig.~\ref{fig:IMspecDialow}. We show also
the angular dependence of the $\Lambda(1405)$ production cross
sections in Fig.~\ref{fig:angprodlow}. Due to the substantial 
contribution from diagram 1, which is the single step process, 
the $\Lambda(1405)$ is produced also at forward angles. 

For further low incident momenta, the spectra can be distorted 
by threshold effects. We show the 
$\pi \Sigma$ invariant mass spectra for the 200 MeV/c incident 
$K^{-}$ in the left panel of Fig.~\ref{fig:IMspecDialow}. As seen 
in Fig.~\ref{fig:IMspecDialow} for the $\pi^{+}\Sigma^{-}$ spectrum,
the signal of the $\Lambda(1405)$ production is distorted 
by the strong contributions around 1440 MeV. 
This peak structure is produced by a threshold effect and 
has nothing to do with any resonances. 
The phase space is given by the neutron momentum 
in the c.m frame of the reaction and the pion momentum in 
the rest frame of $\pi$ and $\Sigma$. 
The maximum invariant mass $M_{\pi\Sigma}$ 
for the 200 MeV/c incident $K^{-}$ is at 1460 MeV.
At the maximum invariant mass, the neutron momentum 
becomes zero, while the pion momentum has a maximum value.
Because these two factors compete below the threshold,
the phase space suppression is slow and shows up only 
at the invariant mass very close to the threshold.
This creates the peak structure. 
For the incident momentum of 200 MeV/c, diagram 1 gives the dominant 
contributions, as seen in Fig.~\ref{fig:IMspecDialow}. 
This has as a consequence that most of the $\Lambda(1405)$ is produced
at forward angles (See Fig.~\ref{fig:angprodlow}). 

For incident $K^{-}$ momenta smaller than 200 MeV/c, 
the maximum invariant mass of $\pi \Sigma$
in the $K^{-}d \to \pi \Sigma n$ reaction approaches that of the $\Lambda(1405)$  
and the signal for this resonance becomes unclear. Thus, it is hard
to investigate the $\Lambda(1405)$
properties in this reaction with incident $K^{-}$ momenta smaller than 200 MeV/c.

\begin{figure}
\centerline{\includegraphics[width=9cm]{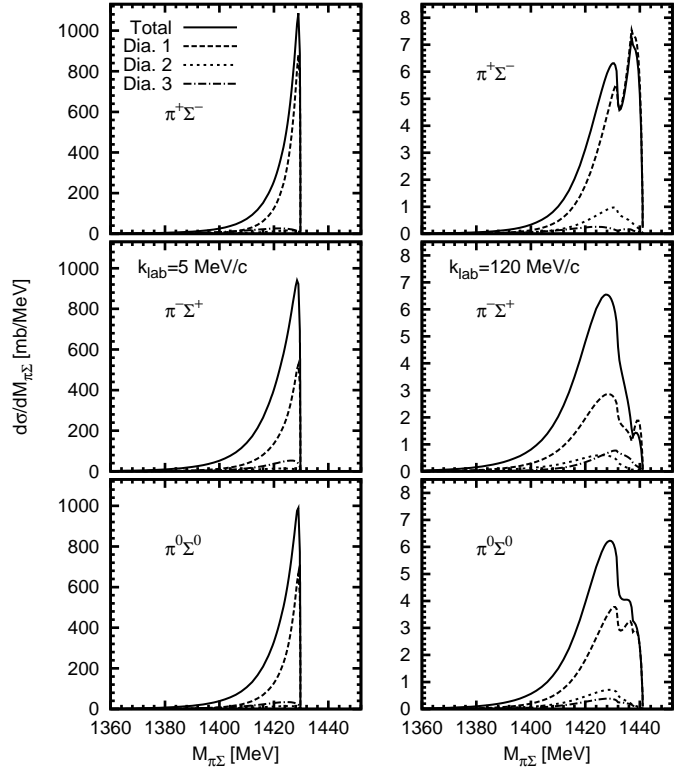}}
\caption{$\pi \Sigma$ invariant mass spectra of $K^{-}d \rightarrow \pi\Sigma n$ 
separately plotted in each diagram contribution.
The plots in the left and right panels 
are calculated with 5 and 120 MeV/c of $K^{-}$ incident momenta,
respectively. Same as described in the caption to 
Fig.~\ref{fig:IMspecDia}. \label{fig:IMspecDialow2}}
\end{figure}

It is important to discuss two particular cases where the experiment 
is likely to be suggested. One of them is the case of kaons at rest, 
which are often used in many other experiments. The other one is the 
case of kaons coming from $\phi$ decay, which are used at FINUDA 
in the Frascati facility.  We have performed the calculations for these 
two cases, in the first one using kaons of 5 MeV/c instead of stopped 
kaons, and in the second case using kaons of 120 MeV/c. The results 
can be seen in Fig.~\ref{fig:IMspecDialow2}. For the low momentum 
case we observe the dominance of the impulse approximation term, 
giving a large contribution close to threshold which extends below 
threshold due to Fermi motion. The important thing to see is that the 
impulse approximation term is absolutely dominant and the trace of 
the $\Lambda(1405)$ is lost in the figure.  One can certainly conclude 
that the case of stopped kaons does not provide a good set up to 
learn about the $\Lambda(1405)$.  On the other hand the situation 
for the Frascati Laboratory is also not very promissing. Although a 
signal of the $\Lambda(1405)$ can be seen, the contribution of 
the impulse approximation term is too large and distorts the signal 
of the resonance which is more clearly seen at higher energies.


\subsubsection{Validity of the approximation of fixing the energy of
the $T_{K^{-}N\to \bar KN}$ amplitude in Eqs.~(\ref{eq:T2}) and (\ref{eq:T3}).}
\label{sec:approx}
In the present work, we have calculated the $\Lambda(1405)$ spectra
under the assumption that the amplitude $T_{K^{-} N \to \bar KN}$
in the double scattering processes, diagram 2 and 3 in Fig.~\ref{fig2},
is factorized from the momentum integral by fixing the invariant energy of 
the scattering amplitude as Eq.~(\ref{eq:W}). This may be a good 
approximation if the amplitude $T_{K^{-} N \to \bar KN}$ does not 
change so much in the relevant momentum range of the integral, 
which is determined by the deuteron wavefunction. Inclusion of 
the integral momentum dependence to the scattering amplitude
is expected not to affect the shape of the $\pi \Sigma$ invariant 
mass spectrum, since the spectral shape is determined by the 
$\bar KN \to \pi \Sigma $ scattering amplitude and its energy 
is fixed by the final state.  Here we show a calculation performed 
with the integral momentum dependence in the amplitude 
$T_{K^{-} N \to \bar KN}$ to see the validity of the approximation. 

\begin{figure}
\centerline{\includegraphics[width=7.5cm]{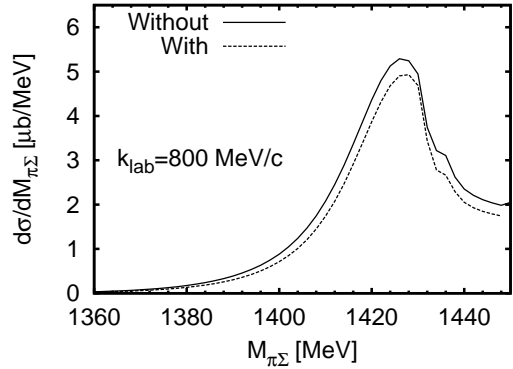}}
\caption{
$\pi \Sigma$ invariant mass spectra of $K^{-} d \to  \pi^{+}\Sigma^{-} n$ 
with the 800 MeV/c incident $K^{-}$
calculated without and with the integral momentum dependence 
of the $T_{K^{-} N \to \bar KN}$ amplitude in the double scattering
diagrams. The solid line denotes the calculation without the momentum
dependence of the ${K^{-} N \to \bar KN}$ amplitude, which is the
same as in the previous calculations. The dotted line is the spectrum obtained 
by the inclusion of the momentum dependence. 
\label{fig:CheckT1}}
\end{figure}

In Fig.~\ref{fig:CheckT1}, we show the $\pi \Sigma$ 
invariant mass spectra of $K^{-} d \to \pi^{+}\Sigma^{-} n$ 
with the 800 MeV/c incident $K^{-}$
calculated without and with the integral momentum dependence 
of the $T_{K^{-} N \to \bar KN}$ amplitude in the double scattering
diagrams. The calculation (the solid line) without the momentum
dependence of the ${K^{-} N \to \bar KN}$ amplitude is the
same as in the previous calculations. The momentum dependence 
on the energy of the $\bar KN \to \pi \Sigma $ scattering amplitude 
has been included by 
\begin{equation}
 W_{1}=\sqrt{(k^{0}+M_{N})^{2}-(\vec q + \vec p_{n})^{2}}
\end{equation}
The dotted line of Fig.~\ref{fig:CheckT1} denotes the spectrum obtained 
by the inclusion of the momentum dependence. 
As seen in Fig.~\ref{fig:CheckT1}, the difference between the two lines is 
small and the spectral shapes are very similar. The strength of the spectra
is slightly different. This is because, for the 800 MeV/c incident $K^{-}$, 
$W_{1}$ can be 1700 MeV with the nucleon in the deuteron at rest,
and the scattering amplitude $T_{K^{-} N \to \bar KN}$ has small 
energy dependence around this energy as seen in Fig.~\ref{fig:ScattAmp}.
Thus, our conclusion that the $\pi \Sigma$ mass spectra have a peak for 
the $\Lambda(1405)$ around 1420 MeV in the $K^{-}d \to \pi \Sigma n$
reaction does not change at all even if we include the momentum dependence 
into the ${K^{-} N \to \bar KN}$ amplitude.

For the absolute value of the production of the $\Lambda(1405)$
in the present reaction, we could have 5\% to 25 \% corrections 
depending on the $K^{-}$ incident momentum from the approximation of the 
loop momentum integral. For instance, in case of the 750 MeV/c incident $K^{-}$, $W_{1}$ can be around 1650 MeV where the $K^{-} p \to 
\bar K^{0} n$ amplitude has a bump caused by the presence of the 
$\Lambda(1670)$ resonance (see Fig.~\ref{fig:ScattAmp}).
Due to the strong energy dependence of the $K^{-} p \to \bar K^{0} n$
amplitude around $W_{1}\simeq 1650$ MeV, we found the correction to be 
25 \% in the absolute value. 
The main consequence of this correction is that the bump structure seen in 
the $\Lambda(1405)$ cross section around $k_{\rm lab} \sim 750$ MeV/c 
in Fig.~\ref{fig:ProdCSLambda} is smeared out and the energy 
dependence of the integrated cross section is now smoother.


\section{Summary}
\label{sec:summary}

We have studied the $K^{-}$ induced production of $\Lambda(1405)$
with a deuteron target by calculating the observable  $K^{-} d \to \pi \Sigma n$
reaction. We have found that, in the $K^{-} d \to \pi \Sigma n$ process,
the $\Lambda(1405)$ resonance is produced by the $\bar KN$ channel,
and, therefore, this process is suited to investigate 
the  properties of the $\Lambda(1405)$ induced by subthreshold $\bar KN$. 

We have calculated the $\pi \Sigma$ invariant mass spectra
of the $K^{-} d \to \pi \Sigma n$ reaction using the $\bar KN \to \bar KN$
and $\bar KN \to \pi \Sigma$ amplitudes obtained by the chiral unitary 
approach. The present calculation agrees with the observed 
spectrum of $K^{-} d \to \pi^{+} \Sigma^{-} n$ in the bubble chamber 
experiment, in which the $\Lambda(1405)$ resonance  appears 
at 1420 MeV not 1405 MeV. The present model also
reproduces the $\Lambda(1405)$ production 
cross section. 
We have also found that the $\Lambda(1405)$ production 
has a strong angular dependence in this reaction and that
the $\Lambda(1405)$ is produced mostly in backward directions
against the incident $K^{-}$. 
We have shown that the production cross section decreases 
as the incident $K^{-}$ momentum increases. 
Due to the worse momentum matching between the kaon and 
the nucleon inside the deuteron, the more energetic incident $K^{-}$s
produce less $\Lambda(1405)$s. 
These findings should be a guideline for future experiments that would help
understand the dynamics and repercussions of subthreshold $\bar K N$ 
production of the $\Lambda(1405)$.

\section*{Acknowledgements}
%
This work was partly supported by
the Grant-in-Aid for Scientific Research
from MEXT and JSPS (Nos.
   20028004    	
   and 20540273), 
the collaboration agreement between the JSPS of Japan and the CSIC of Spain,  
and the Grant-in-Aid for the Global COE Program 
"The Next Generation of Physics, Spun from Universality and Emergence" 
from MEXT of Japan.
This work was also supported in part by DGICYT contract number
FIS2006-03438. 
We acknowledge the support of the European Community-Research
Infrastructure Integrating Activity
``Study of Strongly Interacting Matter" 
(acronym HadronPhysics2, Grant Agreement n. 227431) under 
the Seventh Framework Programme of EU.
Work supported in part by DFG (SFB/TR 16, ``Subnuclear Structure of Matter'').
This work was done under
the Yukawa International Program for Quark-hadron Sciences (YIPQS).

%
%

\end{document}